\documentclass[final]{aipproc}
\layoutstyle{6x9}

\begin{document}

\title{EBAC-DCC Analysis of World Data of $\pi N$, $\gamma N$, and $N(e,e')$ Reactions}


\author{H. Kamano}{
address={Department of Physics, Osaka City University, Osaka 558-8585, Japan},
altaddress={Excited Baryon Analysis Center, Thomas Jefferson National Accelerator Facility, Newport New, Virginia 23606, USA } 
}

\author{T.-S. H. Lee}{
address={Physics Division, Argonne National Laboratory, Argonne, Illinois 60439, USA},
altaddress={Excited Baryon Analysis Center, Thomas Jefferson National Accelerator Facility, Newport New, Virginia 23606, USA}
}

\begin{abstract}
The development, results, and prospect of the Dynamical Coupled-Channels
analysis at Excited Baryon Analysis Center (EBAC-DCC) are reported.
\end{abstract}

\maketitle

\section{Introduction}

The Excited Baryon Analysis Center (EBAC) at Jefferson Lab was 
established in the Spring of 2006 for investigating the  nucleon resonances ($N^*$).
In this contribution, we report on the development, results, and prospect of EBAC.

\begin{figure}[b]
\includegraphics[clip,height=.25\textheight]{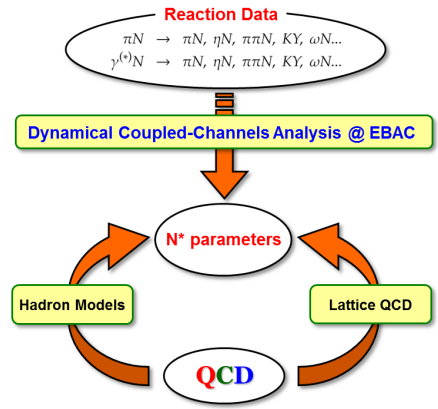}
\caption{The strategy of the EBAC project.}
\label{fig:ebac-strategy}
\end{figure}

The EBAC project has three components, as 
illustrated in Fig.~\ref{fig:ebac-strategy}.
The first task is to perform a  dynamical coupled-channels analysis 
of the {\it world} data of $\pi N, \gamma^*N \rightarrow
\pi N, \eta N, \pi\pi N, K\Lambda, K\Sigma, \omega N,\cdots$
to determine the meson-baryon partial-wave amplitudes. 
The second step is to develop a procedure to extract the $N^*$ parameters 
from the determined partial-wave amplitudes. 
The third step is to investigate the interpretations of the extracted $N^*$ properties 
in terms of the available hadron models and Lattice QCD.

\section{EBAC-DCC model}

The EBAC analysis is based on a Hamiltonian formulation~\cite{msl07} within which 
the reaction amplitudes $T_{\alpha,\beta}(p,p';E)$ in each partial-wave are calculated from
the following coupled-channels integral equations,
\begin{eqnarray}
T_{\alpha,\beta}(p,p';E)&=& V_{\alpha,\beta}(p,p') + \sum_{\gamma}
 \int_{0}^{\infty} q^2 d q  V_{\alpha,\gamma}(p, q )
G_{\gamma}(q ,E)
T_{\gamma,\beta}( q  ,p',E) \,, \label{eq:cct}\\
V_{\alpha,\beta}&=& v_{\alpha,\beta}+
\sum_{N^*}\frac{\Gamma^{\dagger}_{N^*,\alpha} \Gamma_{N^*,\beta}}
{E-M^*} \,,
\label{eq:ccv}
\end{eqnarray}
where $\alpha,\beta,\gamma = \gamma N, \pi N, \eta N, KY, \omega N$, and
$\pi\pi N$ which has  $\pi \Delta, \rho N, \sigma N$ resonant components,
$v_{\alpha,\beta}$ are meson-exchange interactions deduced from
phenomenological Lagrangian, $\Gamma_{N^*,\beta}$  describes
the excitation of the nucleon to a bare $N^*$ state with a mass
$M^*$, and $G_{\gamma}(q ,E)$ is a meson-baryon propagator. 
The EBAC-DCC model, defined by Eqs.~(\ref{eq:cct}) and~(\ref{eq:ccv}), 
satisfies two- and three-body unitarity conditions which are
the most essential theoretical requirements.
Compared with the approaches based on K-matrix or dispersion-relations,
the EBAC-DCC approach has one distinct feature that
the analysis can provide information on reaction mechanisms for interpreting
the extracted nucleon resonances in terms of the coupling of
the bare  $N^*$ states with the meson clouds  generated
by the meson-exchange interaction $v_{\alpha,\beta}$.

\section{Development in 2006-2010}

In order to determine the parameters associated with the strong-interactions
parts of  $V_{\alpha,\beta}$ of Eq.~(\ref{eq:ccv}), the EBAC-DCC model was first applied
to fit the $\pi N$ elastic scattering up to invariant mass $W = 2 $ GeV.
For simplicity, $KY$ and $\omega N$ channels were not included during
this developing stage.
The electromagnetic parts of  $V_{\alpha,\beta}$ were then determined by fitting
the data of $\gamma p \rightarrow \pi^0p, \pi^+n$ and $p(e,e'\pi^{0,+})N$.

The resulting 5-channels model was then tested by comparing 
the predicted $\pi N, \gamma N \rightarrow \pi\pi N$ production cross sections with the data.
In parallel to analyzing the data, a procedure
to analytically continue Eqs.~(\ref{eq:cct}) and~(\ref{eq:ccv}) 
to the complex energy plane was developed to extract
the positions an residues of nucleon resonances.

In the following, we present the sample results from these efforts.

\subsection{Results for single pion production reactions}

In fitting the  $\pi N$ elastic scattering, we found that one or two bare $N^*$ states 
were needed in each partial wave. 
The coupling strengths of the $N^*\rightarrow MB$ vertex interactions
$\Gamma_{N^*,MB}$ with $MB=\pi N, \eta N, \pi\Delta, \rho N, \sigma N$
were then determined in the $\chi^2$-fits to the data.
Our results were given in Ref.~\cite{jlms07}.

\begin{figure}[t]
\includegraphics[clip,height=.35\textheight]{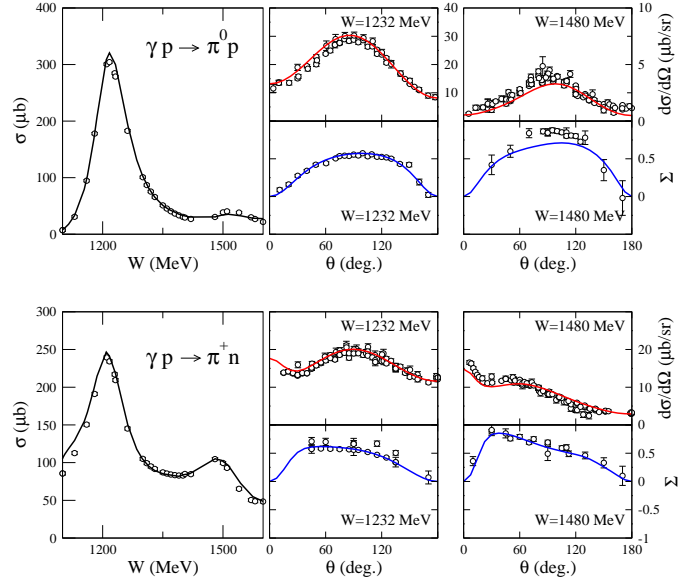}
\caption{The EBAC-DCC results~\cite{jlmss08} of total cross sections ($\sigma$), 
differential cross sections ($d\sigma/d\Omega)$, and photon
asymmetry ($\Sigma$) of $\gamma p\rightarrow \pi^0p$ (upper parts), 
$\gamma p \rightarrow \pi^+n$ (lower parts).}
\label{fig:gnpin}
\end{figure}

Our next step was to determine the bare $\gamma N \rightarrow N^*$
interaction $\Gamma_{N^*,\gamma N}$ by fitting the $\gamma p \rightarrow
\pi^0p$ and $\gamma p \rightarrow \pi^+n$ data. 
We found~\cite{jlmss08} that we were able to fit the data only
up to invariant mass $W = 1.6$ GeV, mainly because we did not adjust 
any parameter which was already fixed in the fits to $\pi N$ elastic scattering.  
Some of our results for total cross sections ($\sigma$), 
differential cross sections ($d\sigma/d\Omega)$, and photon 
asymmetry ($\Sigma$) are shown in Fig.~\ref{fig:gnpin}.

\begin{figure}[t]
\includegraphics[clip,height=.25\textheight]{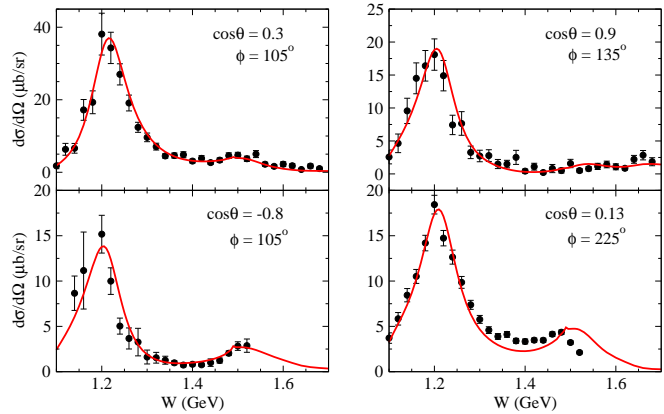}
\caption{The EBAC-DCC results~\cite{jklmss09} of the differential
cross sections of $p(e,e'\pi^0)p$ defined by
$d\sigma/d\Omega = (\Gamma_\gamma)^{-1} d^5\sigma/(dE_{e'}d\Omega_{e'} d\Omega^\ast_\pi)$ 
with $\Gamma_\gamma = [\alpha/(2\pi Q^2)](E_{e'}/E_e)[|\vec q_L|/(1-\varepsilon)]$.
$\theta \equiv \theta^\ast_\pi$ and $\phi \equiv \phi^\ast_\pi$.
}
\label{fig:eepin}
\end{figure}

The $Q^2$-dependence of the $\Gamma_{N^*,\gamma N}$ vertex functions were then
determined~\cite{jklmss09} by fitting the $p(e,e'\pi^0)p$ and $p(e,e'\pi^+)n$ data 
up to $W = 1.6$ GeV and $Q^2 = 1.5 $ (GeV/c)$^2$. Here we also did not adjust
any parameter which was already fixed in the fits to $\pi N$ elastic scattering.
In Fig.~\ref{fig:eepin} we show four of our fits.


\subsection{Results for two-pions production reactions}

\begin{figure}[t]
\includegraphics[clip,height=.15\textheight]{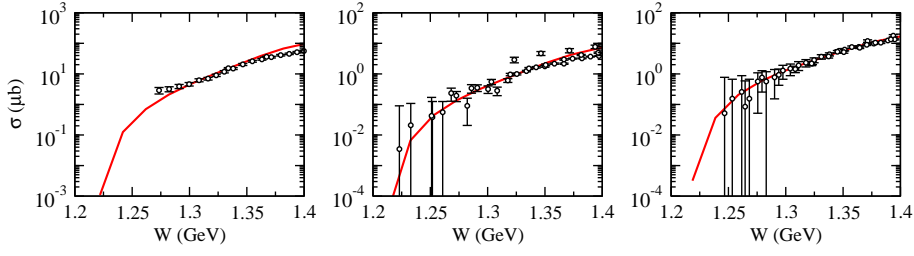}
\caption{The predicted~\cite{kjlms09b} total cross sections of $\gamma p \rightarrow
\pi^+\pi^0 n$ (left), $\pi^0\pi^0 p$ (center), $\pi^+\pi^- p$ (right)
are compared with the data at $W \leq $ 1.4 GeV.}
\label{fig:gnppn}
\end{figure}

\begin{figure}[t]
\includegraphics[clip,height=.35\textheight]{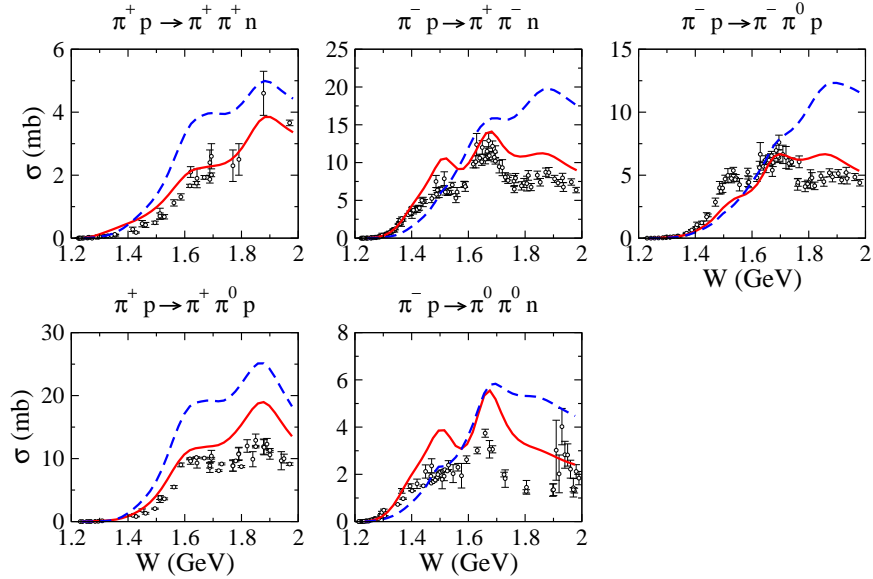}
\caption{The predicted~\cite{kjlms09} total cross sections of
$\pi N \rightarrow \pi\pi N$ are compared with the data.
The dashed-curves are obtained when the coupled-channel effects are turned
off within the EBAC-DCC model.}
\label{fig:pnppn}
\end{figure}

The model constructed from fitting the data of single pion production reactions
was then tested by examining the extent to which the $\pi N \rightarrow \pi\pi N$
and $\gamma N \rightarrow \pi\pi N$ data can be described.
It was found~\cite{kjlms09,kjlms09b} that the predicted total cross sections
are in excellent agreement with the data in the near threshold $W \leq 1.4 $ GeV. 
Our results for $\gamma p \rightarrow \pi^+\pi^- p, \pi^+\pi^0 n, \pi^0\pi^0 p$
are shown in Fig.~\ref{fig:gnppn}. 
In the higher $W$ region, the predicted $\pi N \rightarrow \pi\pi N$ cross sections
can describe to a very large extent the available data, as shown in Fig.~\ref{fig:pnppn}. 
Here the important role of the coupled-channel effects were also demonstrated.
However, the predicted $\gamma p \rightarrow \pi^+\pi^- p, \pi^0\pi^0p$ cross sections
were a factor of about 2 larger than the data while the shapes of
two-particles invariant mass distributions could be described very well.


\subsection{Resonance Extractions}

We follow the earlier works, as reviewed and explained in Refs.~\cite{ssl09,ssl10}, 
to define that the resonances are the eigenstates of 
the Hamiltonian with only outgoing waves of their decay channels.
One then can show that the nucleon resonance positions are
the poles $M_R$ of meson-baryon scattering amplitudes calculated
from Eqs.~(\ref{eq:cct}) and~(\ref{eq:ccv}) on 
the unphysical sheets of complex-$E$ Riemann surface.
The coupling of meson-baryon states with the resonances can be
determined by the residues $R_{N^*,MB}$ at the pole positions.
Our procedures for determining $M_R$ and $R_{N^*,MB}$ and the results
were presented in Refs.~\cite{ssl09,ssl10,sjklms10,knls10}.

\begin{figure}[t]
\includegraphics[clip,height=.2\textheight]{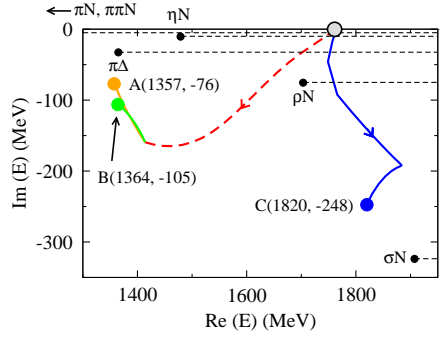}
\caption{The trajectories of the evolution of
three nucleon resonances in $P_{11}$ from the same bare $N^*$ state.
The results were from Ref.~\cite{sjklms10}.}
\label{fig:p11-traj}
\end{figure}
\begin{figure}[t]
\includegraphics[clip,height=.165\textheight]{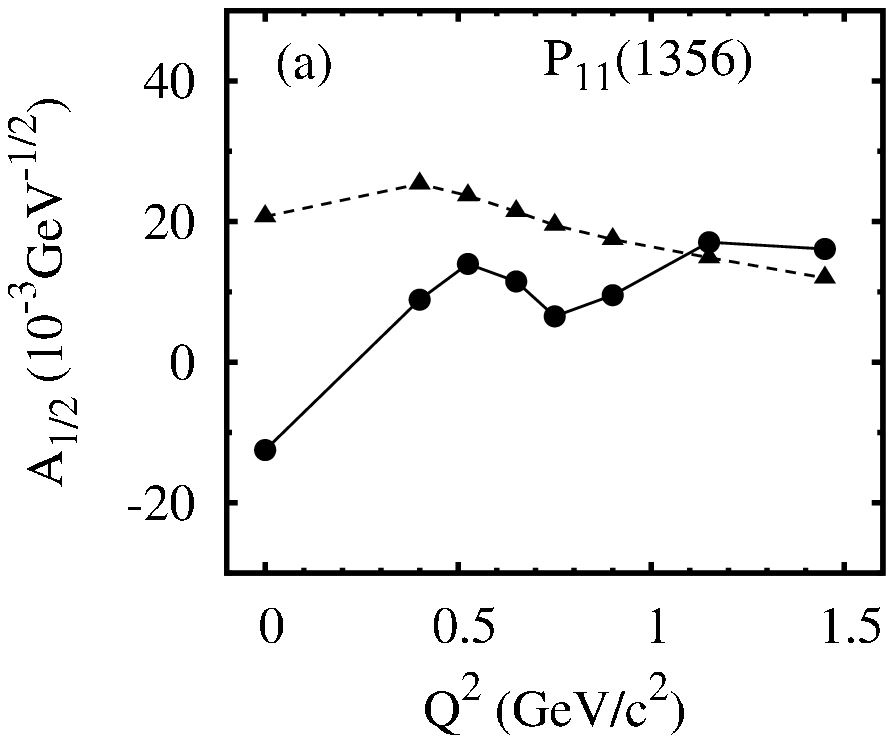}
\includegraphics[clip,height=.165\textheight]{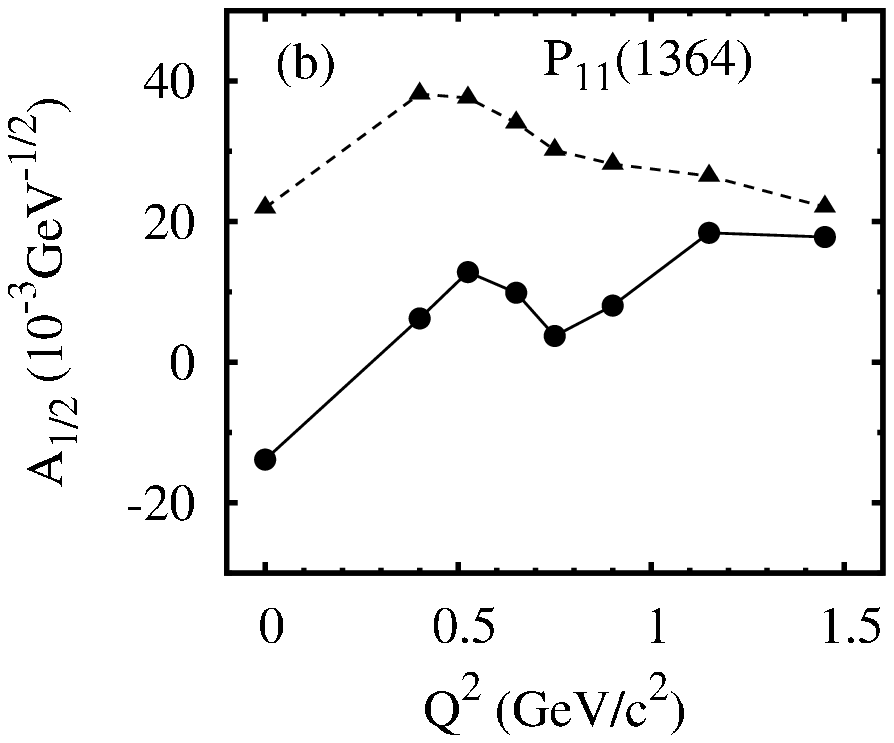}
\includegraphics[clip,height=.165\textheight]{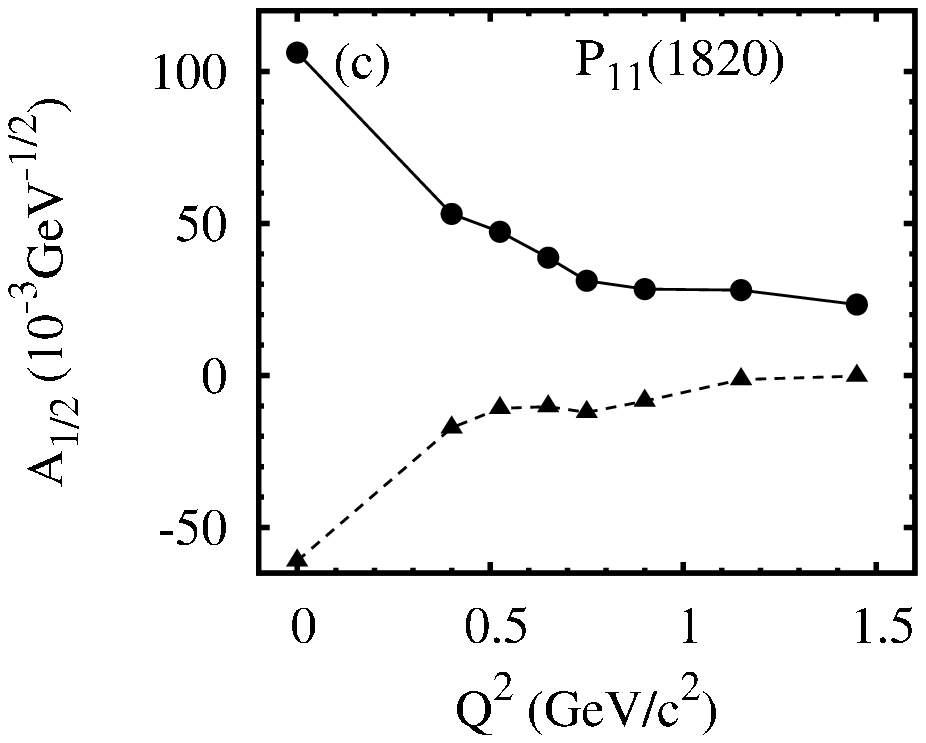}
\caption{The extracted~\cite{ssl10} $\gamma N \rightarrow N^*$ form factors for the
first three $P_{11}$ nucleon resonances. Solid (dashed) curves are their real (imaginary) parts.}
\label{fig:p11-ff}
\end{figure}

With our analytic continuation method~\cite{ssl09,ssl10},
we were able to analyze the dynamical origins of the nucleon resonances
extracted from the EBAC-DCC model.
This was done by examining how the resonance positions move as the
coupled-channels effects are gradually turned off. 
As illustrated in Fig.~\ref{fig:p11-traj} for the $P_{11}$ states,
this exercise revealed that the two poles in Roper region and the next higher
pole are associated with the same bare state.

The extracted residues $R_{N^*,MB}$ are complex which is the necessary mathematical consequences 
of any approach based on a Hamiltonian formulation. 
As an example, the extracted $N^*\rightarrow \gamma N$ form factors for
the three $P_{11}$ resonances indicated in Fig.~\ref{fig:p11-traj}
are shown in Fig.~\ref{fig:p11-ff}.
To complete the EBAC project, we must investigate how
these results can be related to the current hadron models and Lattice QCD.

\section{Prospect}

During the developing stage of EBAC in 2006-2010,  
the EBAC-DCC model parameters were determined by analyzing 
separately the following data: $\pi N\rightarrow \pi N$~\cite{jlms07}, 
$\gamma N \rightarrow \pi N$~\cite{jlmss08}, $N(e,e'\pi)N$~\cite{jklmss09},
$\pi N\rightarrow \pi\pi N$~\cite{kjlms09}, 
and $\gamma N \rightarrow \pi\pi N$~\cite{kjlms09b}. 
The very extensive data of $K\Lambda$ and $K\Sigma$ production were not included in the analysis.
To have a high precision extraction of nucleon resonances, it is necessary to
perform a \emph{combined} simultaneous coupled-channels analysis 
of all meson production reactions.

We have started the first combined analysis of \emph{world} data of
$\pi N, \gamma N \rightarrow \pi N, \eta N, K\Lambda, K\Sigma$ since the summer of 2010. 
Preliminary results have been obtained. 
We expect to complete this task in the Spring of 2012.

The combined analysis must be continued to also fit the $world$ data of meson
electroproduction data for extracting $\gamma N \rightarrow N^*$ form factors 
up to sufficiently high $Q^2$. 
In addition, we should explore the interpretations of the extracted resonance parameters 
in terms of Lattice QCD and the available hadron models,
such as the Dyson-Schwinger-Equation model and constituent quark model.
This last step is needed to complete the EBAC project with conclusive results, 
as indicated in Fig.~\ref{fig:ebac-strategy}.

\begin{theacknowledgments}
The authors thank to B.~Juli\'a-D\'iaz, A.~Matsuyama, S.~X.~Nakamura, T.~Sato, and N.~Suzuki
for their collaboration at EBAC.
TSHL would like to thank A.~W.~Thomas for his strong support to the development of EBAC project
and his many constructive discussions.
This work is supported by the U.S. Department of Energy, Office of Nuclear Physics Division,
under Contract No. DE-AC02-06CH11357,
and Contract No. DE-AC05-06OR23177 under which Jefferson Science Associates operates the Jefferson Lab.
This research used resources of the National Energy Research Scientific Computing Center, 
which is supported by the Office of Science of the U.S. Department of Energy 
under Contract No. DE-AC02-05CH11231, resources provided on ``Fusion,'' 
a 320-node computing cluster operated by the Laboratory Computing Resource Center 
at Argonne National Laboratory, and resources of Barcelona Sucpercomputing Center (BSC/CNS).
\end{theacknowledgments}

\bibliographystyle{aipproc}

\begin{thebibliography}{10}
\expandafter\ifx\csname natexlab\endcsname\relax\def\natexlab#1{#1}\fi
\providecommand{\enquote}[1]{``#1''}
\expandafter\ifx\csname url\endcsname\relax
  \def\url#1{\texttt{#1}}\fi
\expandafter\ifx\csname urlprefix\endcsname\relax\def\urlprefix{URL }\fi
\providecommand{\eprint}[2][]{\url{#2}}

\bibitem[Matsuyama et~al.(2007)]{msl07}
A.~Matsuyama, T.~Sato, and T.-S.~H. Lee, \emph{Phys. Rept.} \textbf{439},
  193--253 (2007).

\bibitem[Juli\'a-D\'iaz et~al.(2007)]{jlms07}
B.~Juli\'a-D\'iaz, T.-S.~H. Lee, A.~Matsuyama, and T.~Sato, \emph{Phys. Rev. C}
  \textbf{76}, 065201 (2007).

\bibitem[Juli\'a-D\'iaz et~al.(2008)]{jlmss08}
B.~Juli\'a-D\'iaz, T.-S.~H. Lee, A.~Matsuyama, T.~Sato, and L.~C. Smith,
  \emph{Phys. Rev. C} \textbf{77}, 045205 (2008).

\bibitem[Juli\'a-D\'iaz et~al.(2009)]{jklmss09}
B.~Juli\'a-D\'iaz, H.~Kamano, T.-S.~H. Lee, A.~Matsuyama, T.~Sato, and
  N.~Suzuki, \emph{Phys. Rev. C} \textbf{80}, 025207 (2009).

\bibitem[Kamano et~al.(2009{\natexlab{a}})]{kjlms09}
H.~Kamano, B.~Juli\'a-D\'iaz, T.-S.~H. Lee, A.~Matsuyama, and T.~Sato,
  \emph{Phys. Rev. C} \textbf{79}, 025206 (2009{\natexlab{a}}).

\bibitem[Kamano et~al.(2009{\natexlab{b}})]{kjlms09b}
H.~Kamano, B.~Juli\'a-D\'iaz, T.-S.~H. Lee, A.~Matsuyama, and T.~Sato,
  \emph{Phys. Rev. C} \textbf{80}, 065203 (2009{\natexlab{b}}).

\bibitem[Suzuki et~al.(2009)]{ssl09}
N.~Suzuki, T.~Sato, and T.-S.~H. Lee, \emph{Phys. Rev. C} \textbf{79}, 025205
  (2009).

\bibitem[Suzuki et~al.(2010{\natexlab{a}})]{ssl10}
N.~Suzuki, T.~Sato, and T.-S.~H. Lee, \emph{Phys. Rev. C} \textbf{82}, 045206
  (2010{\natexlab{a}}).

\bibitem[Suzuki et~al.(2010{\natexlab{b}})]{sjklms10}
N.~Suzuki, B.~Juli\'a-D\'iaz, H.~Kamano, T.-S.~H. Lee, A.~Matsuyama, and
  T.~Sato, \emph{Phys. Rev. Lett.} \textbf{104}, 042302 (2010{\natexlab{b}}).

\bibitem[Kamano et~al.(2010)]{knls10}
H.~Kamano, S.~X. Nakamura, T.-S.~H. Lee, and T.~Sato, \emph{Phys. Rev. C}
  \textbf{81}, 065207 (2010).

\end{thebibliography}

\end{document}